# 'Entanglement' – A new dynamic metric to measure team flow

Gloor, P. A., Zylka, M. P., Fronzetti Colladon, A., & Makai, M.

# 'Entanglement' – a new dynamic metric to measure team flow

We introduce "entanglement", a novel metric to measure how synchronized communication between team members is. This measure calculates the Euclidean distance among team members' social network metrics timeseries. We validate the metric with four case studies. The first case study uses entanglement of 11 medical innovation teams to predict team performance and learning behavior. The second case looks at the e-mail communication of 113 senior executives of an international services firm, predicting employee turnover through lack of entanglement of an employee. The third case analyzes the individual employee performance of 81 managers. The fourth case study predicts performance of 13 customer-dedicated teams at a big international company by comparing entanglement in the e-mail interactions with satisfaction of their customers measured through Net Promoter Score (NPS). While we can only speculate about what is causing the entanglement effect, we find that it is a new and versatile indicator for the analysis of employees' communication, analyzing the hitherto underused temporal dimension of online social networks which could be used as a powerful predictor of employee and team performance, employee turnover, and customer satisfaction.

# 1 Introduction

Albert Einstein called quantum entanglement "spooky action at a distance" (Einstein et al., 1935), predicting that quantum mechanics should allow objects to influence each other's action at great distance. It took other Nobel prize winning physicists' decades after Einstein's death to confirm his prediction. In this paper we propose a similar social entanglement effect between people. Note that we are not making any conclusive claim about the cause of this social entanglement effect, we just find that it seems to exist, and posit that there seem to be useful parallels between quantum entanglement and social entanglement that assist in the conceptualization of the latter.

"You share everything with your bestie. Even brain waves." (Angier, 2018). This is how the New York Times summarized the work of Parkinson, Kleinbaum and Wheatley (2018), who found that brain scans of close friends show similar patterns as they watch a series of short videos. Using these results, the researchers trained a computer algorithm to predict the strength of a social bond between two people based on the relative similarity or synchronization of their neural response patterns. Such neural synchronization patterns are also observed in various other studies in different contexts, e.g., to determine neural contingencies between musical performers and their audiences. Hou et al. (2020) assess the neural synchronization between violinist and audience and the relation to popularity of violin performance. Their findings suggest that neural synchronization between the audience and the performer might serve as an underlying mechanism for the positive reception of musical performance. Further, neural synchronization can be confirmed by analyzing verbal group communication (Liu et al., 2019). Individuals try to achieve neural and body synchronization in order to facilitate fluid interaction (Fairhurst et al., 2013; Yun et al., 2012). Experiments show that synchrony of fingertip movement and neural activity between two persons increases after cooperative interaction (Yun et al., 2012). Hence, engaging individuals in synchronized activities like walking, dancing etc. is an effective way of increasing subsequent cooperation between those individuals.

However, the studies mentioned above focus on neural or body synchronization and are not applied in typical work environments or contexts. But "being in sync" or "in flow" in work environments is a relevant research topic and should be considered by decision-makers to determine the impact of such behavior on employee performance. Being in sync with others can increase cooperation by strengthening social attachment among team members (Wiltermuth and Heath, 2009). Thus, it might also affect team productivity and team performance positively. Offline and online communication plays an important role to distinguish between teams that are in sync or "out of sync". Where offline communication like face-to-face meetings establish team synchronization easily (Maznevski and Chudoba, 2000), online communication such as e-mail and chat tools might diminish team synchronization (Hinds and Bailey, 2003). The asynchronous characteristic of online communication, for instance caused by time lags (Cramton, 2001), may hinder developing a shared team rhythm (Hinds et al., 2015; Hinds and Bailey, 2003).

However, there exist opportunities to analyze online communication data in near-real time for continuous monitoring of team learning and performance. Metrics based on communication flow from person to person or amount of communication are suitable for real-time processing. In addition, studies have shown that analyzing online communication data in organizational contexts (de Oliveira et al., 2019; Gloor et al., 2017b) could be used as a predictor for job-related constructs, such as employee turnover or employee performance. Speed of responding to an e-mail, for example, is a good predictor of individual and team performance (Gloor et al., 2020). It might be a proxy for the passion of the person who is responding to an e-mail (Gloor, 2017), or for other external reasons such as urgency, power differentials, etc.

Based on these behavioral and neuroscientific insights and findings on the relationship between interpersonal synchronization and communication, we hypothesize that being in sync can also be shown by analyzing patterns of team online communication gathered through a social network analysis (SNA) approach. Hence, our research questions are:

(1) Are time series of communication patterns from online communication valid indicators for analyzing the synchronization of a team and its flow?
(2) Is a measure for team flow capable to predict job-related outcomes such as job performance or employee turnover?

We answer these questions by introducing a metric called *entanglement*, which measures the synchronization of e-mail communication behaviors of team members and their flow state over time. This metric is grounded in SNA and identifies the similarity of timeseries of SNA metrics. We validate the metric by conducting four case studies, with different datasets from different organizations. Each case study is in a different context and variants of the entanglement measure are used as a predictor of different individual and group performance indicators.

The rest of the paper is organized as follows. In Section 2 we present the theoretical background of flow state and team synchronization. Subsequently, we illustrate the idea of our entanglement metrics, which want to capture how much people interact in the same rhythm or are "in sync". We finalize this section with the metric's formalization. In Section 3, we explain the data collection and applied methods. Then in Section 4, we introduce four case studies in which we demonstrate the predictive power of the proposed entanglement metrics. In the last section, we discuss results and advocate future research.

## 2 Theoretical background

### 2.1 Team synchronization and flow state

Synchronization is a fundamental element of life. Besides neuronal synchronization mentioned in the introduction, one finds studies that deal with the synchronization of human activities (Guastello and Peressini, 2017). Synchronization is often defined as the manifestation of unintended coordination. It is part of the natural behavior of a human being and takes place so invisibly that we usually do not notice it. It is triggered by audio-visual stimuli, haptic perception or simply by the presence of certain people. Synchronization can be analyzed as neuromuscular coordination, where there is a relatively exact or proportional tracking of body, hand and head movements, autonomic arousal, or electroencephalogram (EEG) readings between two or more people (Guastello and Peressini, 2017). For example, Néda et al. (2000) show that the audience of a concert synchronizes its applause after an asynchronous start and Fairhurst et al. (2013) and Yun et al. (2012) show that people synchronize their finger tapping to improve coordination. While these studies only look at synchronization as neuromuscular coordination and task coordination, there are research efforts currently underway to uncover connections between synchronization in cognition, task structures, and performance outcomes in teams (Gipson et al., 2016).

Better work performance outcomes would also be expected when teams are similarly synchronized (Elkins et al., 2009; Stevens et al., 2013).

The hypothesis that team synchronization leads to better performance is further motivated by the theory of *flow state*. While the concept of synchronization in the above-mentioned studies applies a natural science perspective, human sciences like positive psychology consider synchronization as a part of *flow state* (Gloor et al., 2012) and expect flow state to cause better performance. A team is in flow state (Csikszentmihalyi, 1996) when members create a sense of shared confidence and empathy, which culminates in a collective mental state in which individual intentions harmonize and are in-sync with those members of the group. This condition is also referred to as achieving a "group mind", which is marked by a deep emotional resonance which enables e.g., jazz musicians to be completely coordinated throughout the improvisational flow. In other words, group flow manifests itself in physical and verbal activities, for instance people mirroring each other and quickly finishing each other's sentences using the same words and phrases, indicating a "parallel synchronization of thought" (Armstrong, 2008). The more the team members are in-sync, the more likely it is to observe group flow. Group flow can be analyzed applying "interaction analysis", which entails closely observing and categorizing the interactions, movements, and body language of group members. But, it cannot be limited to neurological studies of particular participants of the group's emotional conditions or subjective memories (Sawyer, 2003). Thus, group flow cannot be split down into specific tasks; rather, it is a process that arises from group dynamics and has the ability to improve job satisfaction, intrinsic motivation, vigor, performance or efficiency (Delarue et al., 2008; Sawyer, 2003; van den Hout et al., 2018). Hence, flow represents rather an oscillating dynamic state that combines continuous and sudden changes across time (Ceja and Navarro, 2012) than a static one.

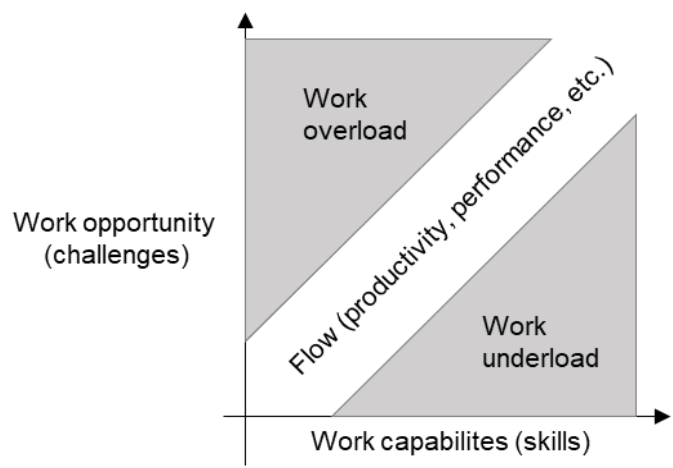

*Figure 1. General model of flow state in work environments. Based on original model of flow state by Csikszentmihalyi (2000).*

The flow concept can be transferred into the organizational context (Heyne et al., 2011). Bakker (2005) defines work-related flow as a short-term peak experience at work that is characterized by absorption, work enjoyment and interest. Teams "are in flow" if there is a certain balance between challenges and the skill sets of the individual team members. Work-related flow leads to a better productivity and performance (see Figure 1). Further, by the definition of flow by Csikszentmihalyi (1996) high flow leads to high performance. If a team is collectively in flow, it therefore will deliver high performance. In general, flow is likely to correlate positively with measurable results (Quinn, 2005). Quinn (2005, p. 611) emphasizes that "[i]n knowledge work […] flow may be a useful concept for understanding performance.". Studies of flow proceed from a broader awareness that team processes like communication need to be studied as events over time (Arrow et al., 2004).

## 2.2 Entanglement conceptualization and formalization

The idea of the entanglement measure is to determine how a person is in sync with his/her group and shares the same flow with the other team members, with regards to communication over a period of time. In an attempt to conceptualize entanglement, a multidisciplinary approach is proposed, bringing together concepts from several disciplines, ranging from quantum mechanics to human and social sciences. The term entanglement is borrowed from quantum physics, where a pair or group of particles which are "entangled" mysteriously change their quantum state at the same time, even when the group of particles is physically far apart at different locations on the world (Horodecki et al., 2009). A result of this phenomenon is that when one measures the quantum state of one particle, one simultaneously determines the quantum state of the other particle. A quantum state (of a particle) is a representation of knowledge or information about an aspect of the system or reality (Pusey et al., 2012). In this study, we interpret the reality as the state about a person-to-person relationship. Thus, the two particles are seen as two individuals that have potentially interacted with "others", not necessarily with each other, and have therefore become entangled. Our idea of synchronicity is that people are in-sync when they show similar behavioral patterns, such as communication activity. Hence, two persons are entangled even when they are physically separated or not involved in a (local) interaction with each other but share a similar communication behavior (an example is provided in Figure 2).

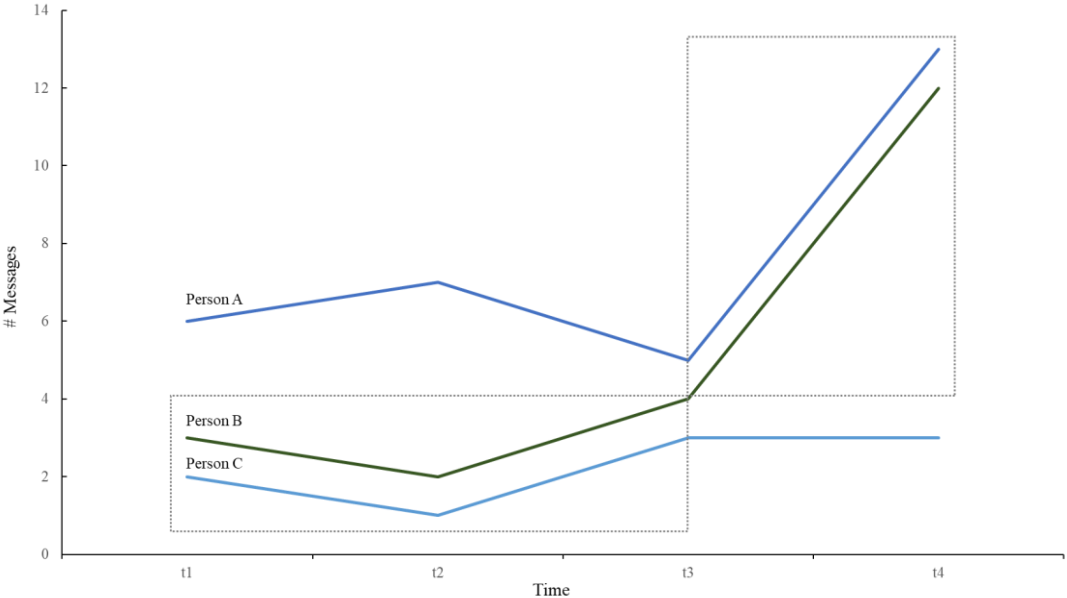

*Figure 2. Communication intensity of three persons by time*

Similar concepts have previously been described in psychology and sociology. "Entrainment" describes a process where one system's motion or oscillation frequency synchronizes with another system, for instance the brainwaves of two people rocking together in their chairs. Cross et al. (2019) define interpersonal entrainment as the synchronization of organisms to a rhythm, for example singing, dancing, or even walking together. Much earlier, early twentieth century French sociologist Emile Durkheim defined collective effervescence as the similar but broader notion of synchronized action between humans (Durkheim, 2008), to describe when a community or society comes together to communicate the same thought or participate in the same action. This concept has been picked up by sociologist Randall Collins through his construct of "Interaction Ritual Chains" (Collins, 2005), which explain collective action through shared emotional energy. The common theme of all these constructs is colocation, people creating and experiencing emotional energy by being together at the same location. We therefore prefer the term "entanglement" to describe synchronous action between humans independent from where they are located, to describe in the words of Albert Einstein, "spooky action at a distance".

Human communication is fundamentally synchronous and rhythmic, two important characteristics of individual and interactional behavior (Condon, 1986). The synchronization of interactional behaviors helps to generate a sense of flow state for the persons involved (Condon, 1986). Further, it always takes other people for a person to reach the state of flow (Collins, 2005), while the other people do not have to be physically present. Thus, entanglement leads to a flow state of two persons analogous to the "mysterious change" of a particle's quantum state. Intuitively, we propose that the "more similar the communication" of two persons A and B is, the more person A is in sync and is able to share the same flow of communication with person B over a period of time. Individuals that are in flow might have higher abilities to productively channel their cooperative spirit when working together.

Figure 2 shows the communication of three persons by time. Person B and C communicate in similar intensity (here: number of sent messages) from *t1* until *t3*. Their communication decreases from *t1* to *t2* and increases from *t2* to *t3* by the same amount. Further, their lines in the chart are very close together meaning the distance between each of their data points is short. We observe the same pattern for person A and person B in time period *t3* and *t4*. Such patterns might indicate synchronization. Thus, we can state that the distance of the data points representing the communication intensity between two or more persons in a specific time window is an indicator for their synchronization. Here, we use the Euclidean distance, a straight-line distance between two points in Euclidean space. We calculate the Euclidean distance *d* of two data points *x* and *y* of a communication metric *A* of the same time window *t* with:

$$d(A(x_t), A(y_t)) = \sqrt{(A(x_t) - A(y_t))^2}$$

This Euclidean distance specified in the formula above is calculated for every pair of nodes and time window *t*.

An essential requirement to determine if persons are entangled is to consider both team synchronization and team flow. Team flow is based on flow experienced in relational embeddedness (Burt, 2005) which can be established by e.g. communication and collaboration. To address this structural feature of communication, we propose to apply SNA. SNA offers a suitable methodology to study group dynamics as well as to investigate the role of the individuals within these dynamics (Wasserman and Faust, 1994). It focuses on various aspects of the relational structures and the flow of information, which characterize a network of people, through graphs and structural measures.

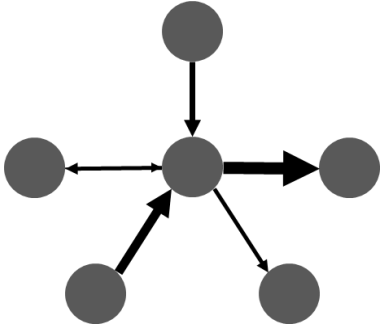

*Figure 3. Graph representing an e-mail communication network*

To better illustrate the concept of "entanglement" we consider an e-mail network, characterized as a graph made of a set of nodes (e-mail accounts) and a set of directed edges (weighted by the number of e-mails) connecting these nodes. The direction of an edge specifies the source (e-mail sender) and target (e-mail receiver) node; the weight of an edge shows the relation intensity (number of e-mails) between two nodes (see Figure 3). For example, if person A sends 3 emails to person B, we see an arc originating at node A and terminating at node B of weight equal to 3.

To illustrate the idea and calculation of entanglement with an example, we use an individual mailbox representing a dataset of e-mails of persons that work together on several projects. First, we collected the mailbox and stored it in a database, where the e-mail data was structured from a network perspective. In order to calculate the entanglement of the mailbox owner and his/her colleagues, we take the inverse of the Euclidean distance of the time series of the communication activity represented by messages sent over time for each node/actor in the network. This value will get the larger the more similar the activity time series of two actors are. However, we have to distinguish between two pairs of actors at different locations in the network, one pair embedded into a tight cluster communicating with many other actors, while the other pair is exchanging the same number of e-mails as the first pair, but is only weakly connected to other actors. To make this metric comparable among pairs of actors with different levels of activity in the same network, we multiply it by the product of the degree centralities of both actors. Degree measures the centrality, sometimes seen as a proxy of popularity, of a node in a network, by counting the number of its nearest neighbors (Freeman, 1978). Further it can be a proxy for the level of engagement within a group, team or organization (Gloor et al., 2020). Communication activity via e-mail (Gloor et al., 2014) indicates the number of e-mail messages sent by a person within a time interval.

Figure 4 shows the e-mail communication activity over a period of time, for the email box we analyzed. The blue line shows the mailbox owner's communication activity, the other lines correspond to the people s/he is exchanging e-mails most frequently with. The more correlated the communication activity between the owner of the mailbox and another person are, the more they are in sync, share the same flow over a period of time, and thus are entangled. The picture also illustrates the need to include degree centrality in the entanglement formula, as the levels of activities, while running in parallel, are vastly different for different people.

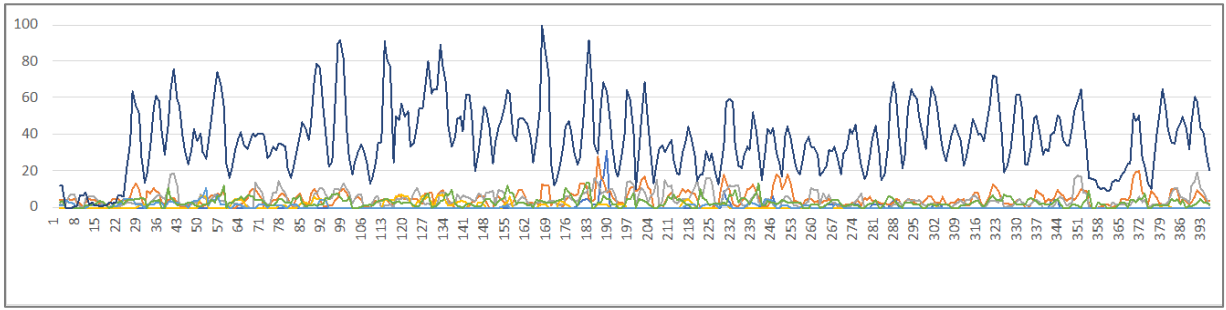

*Figure 4. Flow of e-mail communication by time*

Accordingly, we define the activity entanglement $E_A(x_T, y_T)$ between two individuals, named *x* and *y* in a specific time window *T*, as:

$$E_A(x_T, y_T) = \frac{C_D(x_T) \, C_D(y_T)}{d(A(x_T), A(y_T))}$$

where $C_D(x_T)$ and $C_D(y_T)$ are the degree centralities of the two individuals *x* and *y*, and $d(A(x_T), A(y_T))$ is their Euclidean distance, with respect to communication activity *A* in a defined time window *T*. In other words, the entanglement of two individuals *x* and *y* is given by the multiplication of the number of their direct contacts in the e-mail network divided by their synchronization of communication activity. As has been said above, it is necessary to include the product of the degree centralities of *x* and *y* into the entanglement formula to provide for the differences in centralities between actors: assume that actor *x* has low degree, if *x* is synchronized with highly connected actor y having high degree centrality, the high degree of actor *y* will boost entanglement of actor *x* in comparison with all other actors in the network. In other words, we want our metric to reward less influential actors that are synchronized with influential actors.

Similarly, we could consider not just communication activity, but also individuals' synchronization in weighted and unweighted betweenness centrality. Betweenness is a well-known metric in social network analysis. It is the sum of the fraction of all-pairs shortest paths that pass through a node $v$ (Freeman, 1977):

$$C_B(v) = \sum_{s \neq v \neq t \in V} \frac{\sigma(s,t|v)}{\sigma(s,t)},$$

where $V$ is the set of nodes, $\sigma(s,t)$ is the number of shortest paths from *s* to *t*, and $\sigma(s,t|v)$ is the number of those paths passing through node $v$ (Brandes, 2001). Inverse arc weights are considered for the determination of node distances. To control for network size, the above index is usually normalized between zero and one.

If the betweenness centrality time series of two individuals are in sync, it means that they share similar network positions, and levels of influence, at the same time. Individual betweenness entanglement $E_B$ is the product of the degree of two individuals divided by their Euclidean distance in betweenness centrality over a period of time.

$$E_B(x_T, y_T) = \frac{C_D(x_T)\, C_D(y_T)}{d\big(C_B(x_T), C_B(y_T)\big)}$$

In addition, we speculate on the possibility to evaluate how much an individual is in sync with the aggregated flow of the entire network. As a proxy of the aggregated rhythm of the team we take Freeman's group betweenness centralization, $C_{GB}$ (Freeman, 1978). Group betweenness centralization is the sum of the differences between the betweenness centrality of the most central node, $C_B(v^*)$, and that of all other nodes in the network (Freeman, 1978; Wasserman and Faust, 1994), normalized by its maximum value which is $(G-1)^2(G-2)$ where $G$ is the total number of nodes:

$$C_{GB} = \frac{2 \sum_{i=1}^{G}[C_B(v^*) - C_B(v_i)]}{[(G-1)^2(G-2)]}.$$

This definition of group betweenness centralization is appropriate for this use case, as we compare how entangled an individual node is with all other nodes with regards to betweenness.

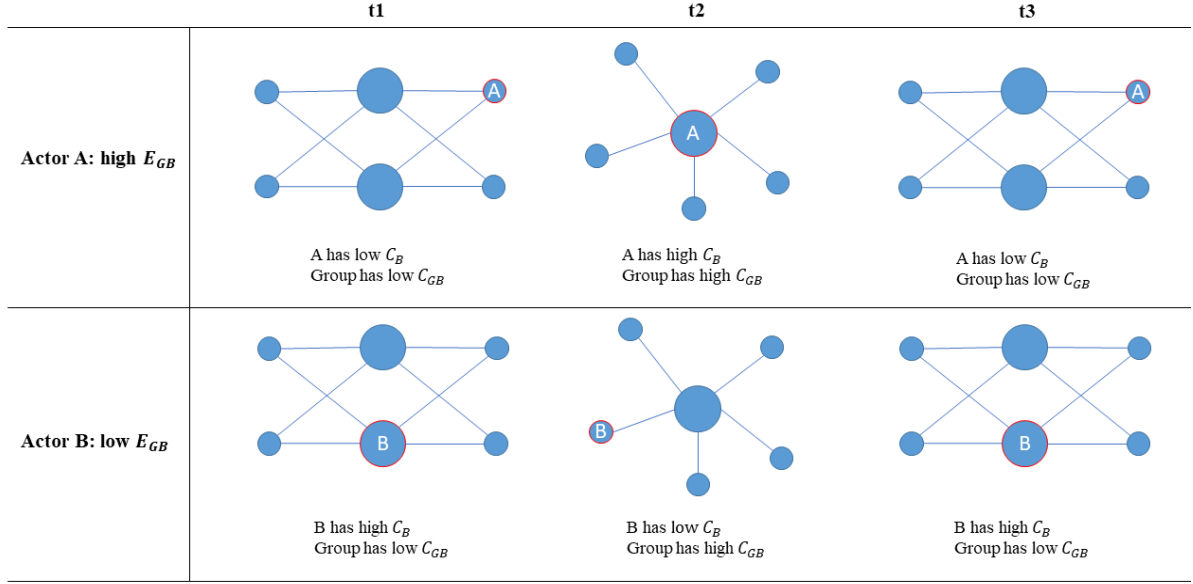

*Figure 5. Intuitive motivation for group betweenness entanglement ($E_{GB}$)*

Figure 5 gives an intuitive motivation for the usefulness of group betweenness entanglement. It shows a group of six actors at three points in time of a changing network structure. Actor A is very much "entangled" with the overall group: In t1 and t3, when the group betweenness centralization ($C_{GB}$) is low, his/her (individual) betweenness centrality ($C_B$) is low also, in t2, when the group betweenness centralization is high, his/her $C_B$ is high too, leading to low Euclidean distance of his/her $C_B$ to $C_{GB}$, resulting in high entanglement. In contrast, actor B is lowly "entangled" with the group, in t1 and t3 when $C_{GB}$ is low, his/her betweenness centrality ($C_B$) is high, in t2 when $C_{GB}$ is high, his $C_B$ is low. This leads to high Euclidean distance to $C_{GB}$, and thus to low entanglement. Formally, we measure group betweenness entanglement $E_{GB}$ by dividing group betweenness centralization $C_{GB}$ by the Euclidean distance of group betweenness centralization and normalized betweenness centrality of the actor being analyzed over a time period. $C_{GB_T}$ – as a metric of variation – is an indicator for the centralization of the group in time window $T$, the individual betweenness centrality $C_B(x_T)$ in this sense is an influence on $C_{GB_T}$, i.e., how much an actor impacts $C_{GB_T}$. Intuitively this metric reflects the contribution of this actor to the level of centralization of its group. In other words, it measures how far away the normalized betweenness centrality of an actor is from the betweenness centralization of its group at any point in time. If an actor's betweenness is high and its group betweenness centralization is high, the actor is probably responsible for the centralized network structure – thus the Euclidean distance between group betweenness centralization and an actor's betweenness centrality is small, and therefore the actor's group betweenness entanglement high. On the other hand, if an actor's betweenness is low and its group betweenness centralization is high, it means somebody else is central and the actor is unimportant in betweenness centrality terms, thus less entangled with the group. We look at this across groups (frequently analyzing advice networks in work settings) and over time. Accordingly, we define group betweenness entanglement, $E_{GB}(x_T)$ of $x$ as:

$$E_{GB}(x_T) = \frac{C_{GB_T}}{d(C_B(x_T), C_{GB_T})}$$

To show the inequality in individual group betweenness entanglement we calculate the Gini coefficient for $E_{GB}$:

$$G(E_{GB}) = \frac{\sum_{i=1}^{n} \sum_{j=1}^{n} |E_{GB}(x_i) - E_{GB}(x_j)|}{2n^2 \overline{E_{GB}}}$$

The same formula can also be used for activity entanglement to calculate $G(E_A)$. Intuitively, the Gini coefficient measures inequality in the distribution of entanglement among all actors in a network. This is based on the observation that for an actor $x$ being resource-poor or resource-rich in a network – the resource being entanglement in this case – can be highly predictive for the behavior or performance of $x$. It therefore makes sense to put the entanglement of $x$ in relationship to the entanglement of all other actors in the network through Gini entanglement.

## 3 Data collection and methods

In this section, we present the data collection process and the methods we applied to analyze the data for the case studies. For each case, we ran the same data collection process. We fetched the e-mails of a sample of project members who chose to participate in each pilot study. All worked at large organizations at the time we collected their communication data. We used Condor[1], a social network and semantic analysis software to collect and analyze the data. We normalized the e-mail data for time zones. In our calculations we set the time window to 7 days, as this has been shown to deliver the best results for this type of organizational e-mail data (Gloor 2017).

We measured the relationship of entanglement calculated from e-mail communication with individual and group outcome variables. Since we explore the properties of communication networks, we focused on the calculation of communication-based measures – such as messages sent and received – and of network centrality measures, as we explained in section 2.2. Further, we used the reach-2 metric, which is the number of nodes that a social actor can reach by going through each of its direct links in the graph (Gloor, 2017). Reach-2 has been used as a proxy for social capital, as it measures the number of connections of the people a person is connected to (de Oliveira and Gloor, 2018).

In addition, we relied on online communication metrics developed specifically for assessing interactivity in e-mail communication. In particular, we looked at the communication activity (Gloor et al., 2014), which indicates the number of e-mail messages sent by a person within a time interval, and at the number of nudges, which represents the average number of pings (emails) that a sender needs to send in order to receive a response from the receiver (Gloor et al., 2014). Here we differentiate between ego nudges (the number of pings before a recipient responds) and alter nudges (the number of pings before others respond). In addition, we measured the contribution index which is the balance between messages sent and messages received (Gloor, 2017). Lastly, we calculated the average response times (ART) to measure how much time it takes a person to reply to an e-mail (Gloor et al., 2014; Merten and Gloor, 2010). This metric is helpful to identify fast and slow communicators and recognize patterns of behavior looking at periods of slower response. We separate between Ego ART, the average number of hours a sender takes to respond to e-mails and Alter ART, the average number of hours recipients takes to respond to a sender.

---

[1] http://www.ickn.org/ckntools.html
Condor is available for free academic use and includes the entanglement metrics.

## 4 Case studies

We illustrate, in four case studies, how the proposed *entanglement* metric can be used with e-mail data to predict work-related outcome variables, such as team performance and employee turnover (see Table 1). The four case studies we present here are related to different business contexts and consider different dependent variables. In all cases we analyze email data, illustrating the suitability of the entanglement metric for online communication data. Our goal here is not to directly compare results across case studies, deriving general conclusions, or claiming causality. Rather we want to show the versatility of our entanglement metrics, which can be adapted to study business interaction dynamics in different scenarios.

| Case study | Industry | Research object | Entanglement measure | Entanglement level | Outcome variable |
|---|---|---|---|---|---|
| A | Health care | 53 employees in 11 healthcare innovation teams | Activity entanglement | Team | Team performance & learning behavior |
| B | Professional services | 113 senior executives | Activity entanglement | Individual | Employee turnover |
| C | Professional services | 81 managers | Betweenness entanglement | Individual | Employee performance |
| D | Professional services | 82 mangers in 13 teams | Group betweenness entanglement | Team | Customer satisfaction |

*Table 1. Case studies overview*

### 4.1 Case study A – learning behavior and performance

This case study was conducted as a pilot in a health care organization to determine if activity entanglement $E_A$ between 53 team members of 11 medical innovation teams could predict performance and learning behaviors.

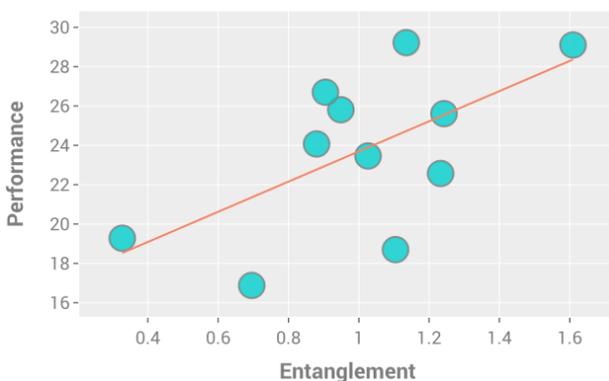
*Figure 6. Entanglement correlation with performance*

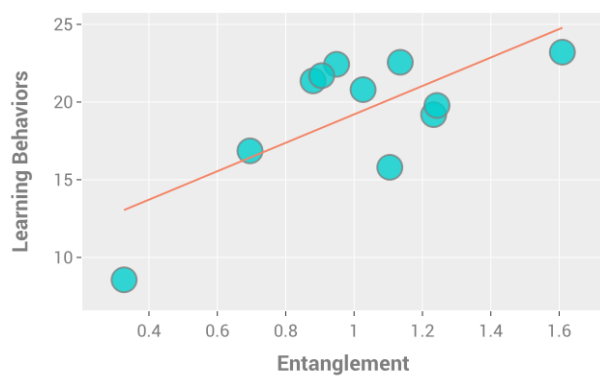
*Figure 7. Entanglement correlation with learning behavior*

The performance and learning behaviors of each team was rated and triangulated every other month for the duration of a year by three overall project managers. They individually rated the team performance and the capability of a team to learn new things. At the same time, all e-mails of the project members were collected and analyzed. Individual activity entanglement of each actor with all other actors was calculated, and then the average was taken for each actor. Finally, for each team average and standard deviation of activity entanglement over all team members was computed.

We find that team performance and learning behavior are significantly correlated with the standard deviation of activity entanglement of team members, as shown in Figure 6 and Figure 7 (which show a scatter plot of the two metrics, with a fitted regression line). The Pearson's correlation coefficient of the standard deviation of activity entanglement of team members with team performance is .615 (p = .045) and with learning behavior is .707 (p = .015). In other words, the wider the spread in activity entanglement $E_A$ of the team members, the higher their performance and learning behavior. This pattern corresponds to a few core team members being strongly entangled, and the remaining members showing weak $E_A$. We also notice that moderate dispersion of entanglement is associated to higher variability in performance scores. This could be explained by control variables we could not collect in this study due to limited data availability. Alternatively, it could suggest that in order for performance to be high, few employees have to take a strong group lead, guiding the others towards a common goal.

## 4.2 Case study B – turnover prediction

In our second case study, we conducted a pilot study at a global professional services firm. In this case we wanted to evaluate the possible association of entanglement with executives' decision to leave the

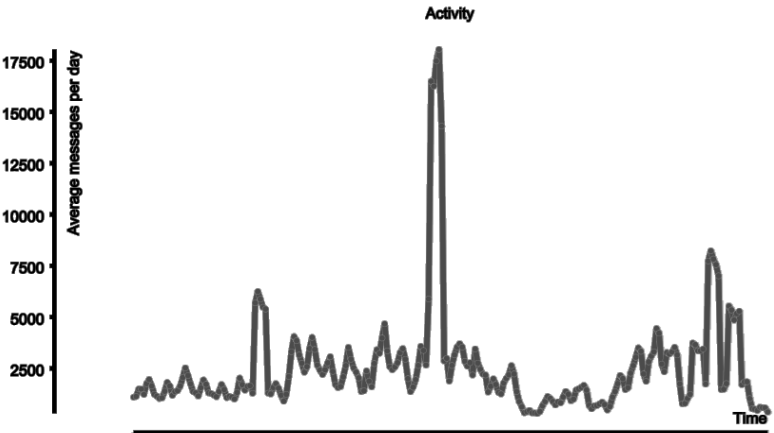
*Figure 8. Communication activity by time*

firm, through voluntary resignation. Turnover of highly important employees such as senior executives is critical for companies, because it has negative implications for firm performance (Hancock et al., 2013; Zylka and Fischbach, 2017).

Eight months of e-mail data of 113 senior executives at a large global services company was collected from May to December 2014 (see Figure 8). We calculated activity entanglement $E_A$ of 55 employees who left the firm from January to May 2015. To determine the inequality in entanglement, we also calculate the Gini index of $E_A$, for each person (from an ego perspective) in the network, considering her/his entanglement and that of all other peers. The Gini index measures the dispersion of entanglement scores of a social actor with all others in the network. In an "egalitarian" network with low Gini index for each node, all actors are either highly or weakly entangled, in a "non-egalitarian" network with high Gini index some actors are highly entangled, while others are weakly entangled. This was compared with the activity entanglement $E_A$ of a control group made of 58 employees, who were selected randomly and still working in an unterminated position at the firm in June 2015.

From a preliminary t-test, we immediately notice that there is a significant difference in the Gini index of activity entanglement, between senior executives who leave the company (M = .457, SD = .070) and those who stay (M = .488, SD = .059), t(111) = -2.513, p = .013. On average, Gini entanglement is significantly higher for those who stay.

Past studies have shown that managerial disengagement might depend on multiple factors and that communication-based and social network analysis metrics, captured from e-mail communication, can reveal it (Gloor et al., 2017b). Accordingly, we present Pearson's correlations (in Table 2) and logistic regression models (in Table 3), to see if the effect of the entanglement variable remained significant when combined with other predictors. The highest correlation of entanglement is with contribution index, which however does not lead to collinearity issues. A high contribution index is an indication for "spammers", the higher the contribution index, the more somebody sends compared to receiving e-mail. If there is a spammer, s/he will be entangled with many, while others who are sending much less, will thus be less entangled. This results in a high Gini entanglement for that person. Extending this effect to all users will lead to high correlation between the two values.

|   |                | 1      | 2       | 3      | 4       | 5       | 6       | 7       | 8       | 9      | 10      | 11      | 12 |
|---|----------------|--------|---------|--------|---------|---------|---------|---------|---------|--------|---------|---------|----|
| 1 | Leaver (1 = yes) | 1    |         |        |         |         |         |         |         |        |         |         |    |
| 2 | Rank           | .056   | 1       |        |         |         |         |         |         |        |         |         |    |
| 3 | Tenure         | .032   | .067    | 1      |         |         |         |         |         |        |         |         |    |
| 4 | TSLP           | -.018  | -.012   | .534** | 1       |         |         |         |         |        |         |         |    |
| 5 | Msg sent       | -.050  | .114    | .196   | -.014   | 1       |         |         |         |        |         |         |    |
| 6 | Msg received   | .040   | .208*   | .129   | -.123   | .632**  | 1       |         |         |        |         |         |    |
| 7 | CI             | -.168  | -.040   | .131   | .039    | .626**  | .219*   | 1       |         |        |         |         |    |
| 8 | Reach 2        | .024   | .306**  | .208*  | -.137   | .431**  | .554**  | .310**  | 1       |        |         |         |    |
| 9 | Betweenness    | .092   | .221*   | .185   | .106    | .445**  | -.018   | .279**  | .236*   | 1      |         |         |    |
| 10| Alter ART      | .071   | -.109   | -.060  | .231*   | -.199*  | -.216*  | -.176   | -.416** | -.060  | 1       |         |    |
| 11| Ego ART        | .233*  | -.216*  | .011   | .066    | -.227*  | -.210*  | -.388** | -.357** | -.066  | .529**  | 1       |    |
| 12| Gini entanglement | -.232* | .014 | .083   | -.044   | .741**  | .554**  | .840**  | .422**  | .208*  | -.243** | -.342** | 1  |

\* p < .05; \*\* p < .01.

*Table 2. Correlations for leavers*

| Variable | Model 1 | Model 2 | Model 3 | Model 4 | Model 5 |
|---|---|---|---|---|---|
| **Rank** | 0.40740 | 0.30430 | 0.06247 | 0.32153 | 0.10214 |
| **Tenure** | 0.00146 | 0.00184 | 0.00094 | -0.00039 | -0.00403 |
| **TSLP** | -0.00080 | -0.00004 | 0.00012 | 0.00036 | 0.00148 |
| **Msg sent** |  | 0.00007 | -0.00036 | -0.00047 | -0.00033 |
| **Msg received** |  | 0.00013 | 0.00049 | 0.00056 | 0.0016226** |
| **CI** |  | -0.91878 | -0.76631 | -0.23275 | 3.326418** |
| **Reach 2** |  |  | -0.00017 | 0.00074 | 0.00037 |
| **Betweenness** |  |  | 0.00004 | 0.00004 | 0.00005 |
| **Alter ART** |  |  |  | -0.00313 | -0.00891 |
| **Ego ART** |  |  |  | 0.021418* | 0.0299733** |
| **Gini entanglement** |  |  |  |  | -35.02065** |
| **Constant** | -0.62053 | -1.12106 | -0.72954 | -1.64071 | 16.28513** |
| **Pseudo R-squared** | 0.00470 | 0.02930 | 0.04970 | 0.08420 | 0.17960 |

*\* p < .05; \*\* p < .01.*

*Table 3. Logistic regression for leavers*

We first tested a model with only the control variables of rank, tenure, and time since last promotion (TSLP) measured in months. In the subsequent models, we added the other predictors in blocks showing, in Model 4, that the only significant predictor, before adding entanglement, is Ego ART. This suggests that managers who leave the company are less responsive to e-mails and take more time to answer. In the full model, Ego ART, messages sent, contribution index and Gini activity entanglement are significant. Including this last predictor in the model leads to a significant improvement of the McFadden's pseudo-R-squared, which more than doubles (going from .08 to .18). As we can see from Model 5, a higher Gini entanglement makes the probability of leaving the company smaller.

To evaluate the possibility of using the entanglement variable for making predictions, we used machine learning. In particular, we used a tree boosting model named CatBoost and its related Python library (Prokhorenkova et al., 2018). This boosting approach is now well-known and proved its usefulness in past research, where it also sometimes outperformed other supervised machine learning methods, such as Support Vector Machines (SVM) and Random Forest Models (Huang et al., 2019). The model performance has been assessed through Monte Carlo Cross Validation (Dubitzky et al., 2007), with 300 random splits of the dataset into train and test data (75% vs 25%). Thanks to the contribution of our variables, we could achieve an average accuracy of predictions of 80.25%, with an average value of the Area Under the ROC-Curve (AUC) of 0.81.

In a second step, we considered the average model resulting from cross-validation and used it to interpret the impact of each variable on predictions (calculated as the average of its absolute Shapley values). We used the *SHapley Additive exPlanations* (*SHAP*) Python package (Lundberg and Lee, 2017). This method proved to be particularly suitable for tree ensembles and to work well also with respect to other approaches (Lundberg et al., 2020, 2018). As Figure 9 shows, the Gini index of activity entanglement is the variable with the highest impact on model predictions. Its contribution is much higher than all other variables, again supporting the importance of this metric. At the second place, we find Ego ART. Results are consistent with those of logit models and indicate that managers who are slower in answering e-mails, and have low Gini entanglement, are more likely to leave the company. Low Gini entanglement means that they show constant levels of entanglement, either being entangled with almost nobody or everyone – a situation that might be stressful to maintain, especially when associated with email overload (Reinke and Chamorro-Premuzic, 2014). Average/high levels of Gini entanglement, on the other hand, have a positive impact on the prediction of staying in the company. This means that these managers show uneven entanglement, being highly entangled with some colleagues, while being weakly entangled with others.

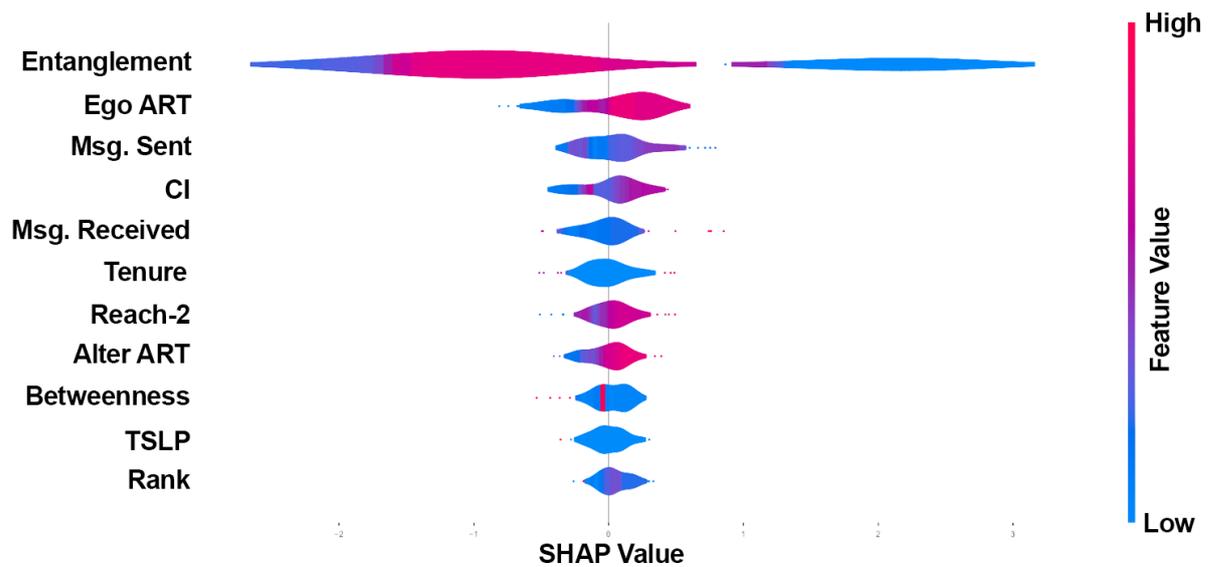

*Figure 9. SHAP values (prediction of leavers)*

### 4.3 Case study C – employee performance

We analyzed the e-mail interactions of 81 managers working for a big international services company. Every year the performance of managers was evaluated by their bosses and by the HR department. Whereas the rating of almost all of these managers was "exceeded expectations" for the year 2015, we noticed that 15 of them obtained a lower rating. Like in the case study B of resigning senior executives, we were interested in understanding if entanglement could be related to individual work performance. Carrying out a t-test, we could see that there is a significant difference between the Gini coefficients of betweenness entanglement $E_B$ scores of top ($M = .508$, $SD = .061$) and low ($M = .469$, $SD = .028$) performers, $t(79) = 2.432$, $p = .017$.

As we did for leavers in case study B, we additionally built logistic regression models to assess the combined impact of variables on the probability to be a low performer. Pearson's correlations among our predictors are presented in Table 4. The highest correlation of entanglement is again with contribution index, but this time lower than case study B.

|   |   | 1 | 2 | 3 | 4 | 5 | 6 | 7 | 8 | 9 | 10 | 11 |
|---|---|---|---|---|---|---|---|---|---|---|----|----|
| 1 | Low Performer (1 = yes) | 1 | | | | | | | | | | |
| 2 | Tenure | .238* | 1 | | | | | | | | | |
| 3 | TSLP | .153 | .178 | 1 | | | | | | | | |
| 4 | Msg Sent | -.070 | .175 | .060 | 1 | | | | | | | |
| 5 | Msg Received | .084 | .080 | .190 | .186 | 1 | | | | | | |
| 6 | CI | -.074 | .206 | .066 | .731** | .180 | 1 | | | | | |
| 7 | Reach 2 | -.038 | .237* | .004 | .200 | .492** | .227* | 1 | | | | |
| 8 | Betweenness | -.082 | .157 | -.039 | .844** | -.031 | .517** | .228* | 1 | | | |
| 9 | Alter ART | -.136 | -.118 | -.180 | .154 | -.171 | .134 | -.264* | .046 | 1 | | |
| 10 | Ego ART | -.139 | -.155 | -.188 | -.024 | -.125 | .021 | -.276* | -.068 | .406** | 1 | |
| 11 | Gini entanglement | -.264* | -.085 | .007 | .484** | .208 | .548** | .234* | .334** | -.024 | .118 | 1 |

\* p < .05; \*\* p < .01.

*Table 4. Correlations for low performers*

| Variable | Model 1 | Model 2 | Model 3 | Model 4 | Model 5 |
|---|---|---|---|---|---|
| **Tenure** | 0.0105286* | 0.011992** | 0.0138258** | 0.0134347** | 0.0128212* |
| **TSLP** | 0.01189 | 0.01115 | 0.01266 | 0.01019 | 0.00714 |
| **Msg Sent** | | -0.00030 | -0.00166 | -0.00170 | -0.00196 |
| **Msg Received** | | 0.00036 | 0.0017504* | 0.0017746* | 0.0020867* |
| **CI** | | -0.44577 | 1.60707 | 1.69432 | 2.958886* |
| **Reach 2** | | | -0.00006 | -0.00088 | 0.00144 |
| **Betweenness** | | | -0.0104667** | -0.0101837** | -0.0085401* |
| **Alter ART** | | | | -0.00636 | -0.01227 |
| **Ego ART** | | | | -0.00733 | 0.00157 |
| **Gini entanglement** | | | | | -26.67039* |
| **Constant** | -3.0535**** | -3.608166*** | -2.26197 | -1.44970 | 10.88953 |
| **Pseudo R-squared** | 0.06980 | 0.09900 | 0.22400 | 0.23140 | 0.28030 |

*\* p < .10; \*\* p < .05; \*\*\* p < .01; \*\*\*\* p < .001.*

*Table 5. Logistic regression for low performers*

As

Table 5 shows, in the full model the p-value of Gini entanglement is only < .1; however, the inclusion of this variable leads to a good improvement of the McFadden's pseudo-R-squared, from .2314 (Model 4) to .2803 (Model 5). A significant performance improvement is also obtained by including weighted betweenness centrality.

The usefulness of the entanglement predictor is confirmed by the results of the CatBoost model that we trained to classify managers into top and low performers. We followed the same procedure as in the previous case study B – i.e., a Monte Carlo cross-validation with 300 repetitions – and obtained good average results (Accuracy = 74,73%, AUC = 0.68). Figure 10 shows the Shapley values associated with each predictor. For an easier reading, we coded top performers as 1 and low performers as 0 (here the model is predicting top performers, which is exactly symmetrical to the choice of predicting low performers that we did in Table 5). Tenure, betweenness centrality and entanglement are the most important predictors – with high Gini coefficient of betweenness entanglement and high betweenness centrality significantly increasing the chance of being classified as a top performer. These managers are highly entangled with some colleagues, and weakly entangled with others – demonstrating selective communication behavior with close collaborators, while being efficient with their time and communicating comparatively less with the rest of the organization. Regarding tenure, we observe the opposite effect, with recently hired

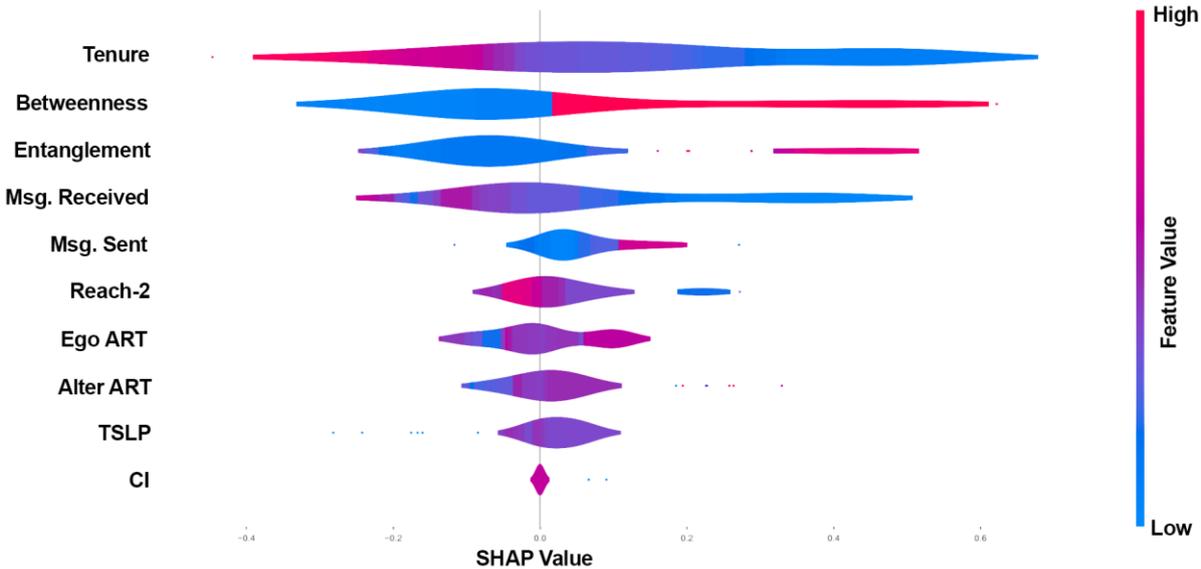

employees generally receiving better ratings.

*Figure 10. SHAP values (prediction of top performers)*

## 4.4 Case study D - customer satisfaction

In this case study, we show that entanglement is significantly related to team performance, measured as customer satisfaction through the Net Promoter Score (NPS). 13 teams within the company participated to our study, comprising a total of 82 managers. Each team was dedicated to a specific client.

We measured betweenness entanglement of each team by taking the group betweenness entanglement of each member and considering group dispersion by means of the Gini coefficient.

We find that high group betweenness entanglement inequality is positively related to team performance – this time measured as customer satisfaction. Running a Pearson's correlation test, we find a significant association of Gini group betweenness entanglement with team performance ($r = .522$, $p = .002$). For each team we have repeated measures over three time periods. Therefore, we used multilevel linear models (Hoffman and Rovine, 2007; Nezlek, 2008; Singer and Willett, 2009) as a more appropriate technique to evaluate the possible effect of entanglement on customer satisfaction. We nested repeated measures into groups (level 2). Results are presented in Table 6.

|  | Model 1 | Model 2 |
|---|---|---|
| Gini Group Betweenness Entanglement |  | 0.6418315* |
| Constant | 0.5244776*** | 0.2869091* |
| Variance L2 | 0.0655537 | 0.0455178 |
| Variance L1 | 0.020654 | 0.0211838 |
| Variance Change L2 |  | -30.56% |
| Variance Change L1 |  | 2.57% |

*Note.* $* p < .05$; $*** p < .001$.

*Table 6. Multilevel models for customer satisfaction (N = 34, with 13 groups)*

As the table shows, the biggest variance proportion can be attributed to team characteristics: the intraclass correlation coefficient is 0.7604, meaning that 76% of the empty model variance is at level 2 (Model 1). Including the entanglement variable in the model (Model 2) reduces this variance of 30.56%, which is a highly significant result for a single predictor. The higher the inequality in group betweenness entanglement is, the happier the customer is. Similarly to case study A, this confirms that selective communication of teams, where some team members are highly entangled and others are not, leads to happier customers.

# 5 Discussion and conclusions

In this study, we propose a novel synchronization metric, called *entanglement*, which is based on SNA of e-mail communication between different actors. We demonstrate with four case studies on real-world datasets that this metric and its variants are a good predictor of different individual and team performance indicators (a summary of our results is provided in Table 7).

| Case study | Dependent Variable | Result summary |
|---|---|---|
| A | Team performance and learning behavior | The wider the spread in activity entanglement of the team members, the higher the team performance and learning behavior. This corresponds to having some core team members strongly entangled and the remaining members weakly entangled. |
| B | Employee turnover | The Gini index of activity entanglement is the variable with the highest impact on model predictions. Employees who stay in the company have high Gini entanglement probably using selective communication and interacting more with some colleagues than with all others. They are also more responsive to emails and take less time to answer. |
| C | Individual performance | Tenure, betweenness centrality and Gini entanglement are the most important predictors of top performers. – with high Gini index of betweenness entanglement and high betweenness centrality significantly increasing the chance of being classified as a top performer. |
| D | Customer satisfaction | The Gini index of group betweenness entanglement for teams, is related to customer satisfaction. The higher the inequality in group betweenness entanglement is for a team, the happier its customer is. This suggests that customers are happier when a few entangled leaders emerge in the team. |

*Table 7. Case study results summary*

Firstly, we find that dispersion of activity entanglement is positively associated with team performance. This means that the synchronized communication activity of some team members and their continuous similar flow state improve the performance of the team. These findings resemble studies showing that e-mail communication and face-to-face communication frequency (Patrashkova-Volzdoska et al., 2003), and flow in knowledge work (Quinn, 2005), can both lead to higher team performance. It also seems that the best teams exhibit higher dispersion, comprising highly entangled team members and more peripheral ones. Teams might benefit from strong leadership of few selected individuals that can guide and inspire others.

With regard to employees disengagement, other studies have already shown that communication-based metrics of SNA can support the prediction of voluntary turnover (de Oliveira et al., 2019; Gloor et al., 2017b). We have proven that our proposed metric entanglement can also predict individual employee turnover and might help such studies to improve their model quality.

Secondly, we show that the Gini coefficient of betweenness entanglement, as well as betweenness centrality, are associated with individual employee performance. A high Gini index of betweenness entanglement significantly increases the chance of being a top performer. This means that focused communication – communicating intensively and highly synchronized with a few select colleagues, while reducing communication with the rest of the organization – is an indicator of high performance. Our findings are consistent with past research (Brass, 1984; Mehra et al., 2001; Sparrowe et al., 2001) showing that network centrality is positively related to individual performance. However, the important part of our metric is that synchronization with others has a positive impact on individual performance, and not only having central social position. Centrality alone may not be enough to explain individual performance (Reinholt et al., 2011) and we address this issue with the betweenness entanglement metric. Furtherly, we found that low tenure also has a positive influence on individual performance.

Thirdly, inequality of group betweenness entanglement in teams positively influences customer satisfaction. The company in case study D considers customer satisfaction as a proxy for team performance. Our findings suggest that the stronger leaders with high entanglement emerge in groups, the happier the customer is. This means we have strongly entangled leaders who influence team dynamics over time, while the rest of the team is rather passive. While Mukherjee (2016) reveals a positive relationship between centralized leadership and sport teams' performance, Mehra et al. (2006) suggests that distributed leadership structures can differ with regard to important structural characteristics, and these differences can have positive or negative effects (Cummings and Cross, 2003) on team performance.

This study contributes to research and practice. First, we contribute to synchronization and flow state research by providing a novel metric for determining communication synchronization in working environments. We validated this metric through four case studies in different business contexts. Flow state research can use our metrics to measure team flow not only by conducting surveys but also by using SNA and taking communication into account. Further, we contribute to human resources (HR) research in providing a novel metric for analyzing employee communication and behaviors. Employee communication is a critical factor for good collaboration and employee and team performance (Gloor et al., 2020; Wen et al., 2019), thus using a new metric of communication dynamics might open new research opportunities. On the other hand, decision makers, such as HR managers, could act as interventionist (Valente, 2012) and use this metric to identify weak and strong entangled actors in the communication network of a team or in the entire company. Thus, HR managers might use this metric to improve performance appraisal systems, anticipate disengagement and improve hiring and retention strategies. Combining novel metrics of e-mail communication analysis with long-established methods to assess employees' satisfaction (like surveys), HR managers can offer improved organizational initiatives, such as mentoring programs or cross-staffing, or retention strategies. The entanglement metric described in this paper has the potential to help managers to better understand the nature of employee online communication at their particular organization. This might lead to a rethinking of team design and building in the specific organization, which could ultimately lead to improved communication and collaboration and might support the identification of cohesive groups.

Nevertheless, e-mail communication analysis combined with SNA raises some ethical concerns. HR managers need to make sure that metrics gathered from such analysis are seen as a support for HR decision making, and not as the holy grail for automated decision making (ADM), without questioning the analysis results. False positives or false negatives can occur and emphasize the supportive character of our metric for HR decisions. The goal of our analytical approach is to support general improvement of group performance and employees' wellbeing, also through the recovery of low-performers. There is potential value for senior leaders in monitoring aggregate behaviors, to understand if there are possible waves of disengagement and address them early at the organizational (if not individual) level. In addition, the entanglement metric offers an opportunity for virtual mirroring sessions (Gloor et al., 2017a), where groups and individuals have the chance to self-reflect on their virtual interactions and communication styles. Through virtual mirroring, employees become aware of their online behavior and this usually triggers change, leading to an improvement in communication and performance ("A Novel Way to Boost Client Satisfaction," 2019).

Future studies might also take body measures like heart rate or body movement into account to determine synchronization and flow state during real-time communication (e.g., online/offline team meeting).

Nowadays, it is technically easy to collect such data e.g. via smart wearables (Gloor et al., 2018). However, we are aware of the difficulty to use such smart devices in an organizational setting, because of security, privacy, and legal issues. Besides e-mail, employees increasingly use instant messaging tools like Slack or Microsoft Teams. Such tools provide application programming interfaces (APIs) for accessing communication data. Researchers could use that data to build a communication network and follow the analytical approach we presented. In general, our conceptualization of entanglement could be extended to other network measures – such as group betweenness centrality as formalized by Everett and Borgatti (Everett and Borgatti, 1999) – or to other aspects of social interaction – such as measuring synchronicity in the emotions of people who carry out similar activities. In future research, we additionally plan to compare our results with those of other possible approaches to the study of temporal networks (Falzon et al., 2018; Holme and Saramäki, 2012).

Our study has some limitations that should be taken into account. While the evidence supports guidance for new research agendas, our analysis is limited to the contexts of the case studies and the available datasets. It will be important to replicate our analysis in organizations of different industries, also considering different job descriptions and hierarchical positions of employees. We included the entanglement calculation in the SNA tools Condor and Griffin, which are free to use for academics in order to facilitate replicability. Further, other social network metrics could be considered to extend our definition of entanglement. For example, additional interaction patterns could be taken into account, developing metrics that specifically look at who communicates with whom. This could be particularly relevant when additional information about nodes is available, other than the social network structure. In addition, we advocate future studies to more deeply investigate the relationship of entanglement with other social network metrics, both time-variant and time-invariant.

Building upon existing synchronization and flow state literature from different disciplines, we showed that the idea of synchronization and flow state can be used together to develop new metrics – based on methods and tools of SNA. Note that social entanglement is an indicator of behavior, with no definitive claims about cause and causality. Just as with quantum entanglement, much more research will be needed to fully "untangle" the origin of social entanglement. Nevertheless, the findings from our four case studies give evidence to the potential of our proposed entanglement metric. We position our research as a starting point for further HR-related analyses, which consider employees' social interactions and communication, with the goal to improve and optimize collaboration, leading to more satisfied employees and customers.